\documentclass[10pt,twocolumn,floatfix,prb,superscriptaddress,]{revtex4-2}
\usepackage{amsmath,amssymb,amsthm,mathrsfs,amsfonts,dsfont,amstext,bbold} 
\usepackage{textcomp,pbox}
\usepackage[export]{adjustbox}
\usepackage{bm}
\usepackage{dcolumn,booktabs,url}
\usepackage[scaled]{helvet}
\usepackage{sansmath,gensymb}
\usepackage{tikz,graphicx,transparent,color}
\usepackage{multirow}
\usepackage[separate-uncertainty = true]{siunitx}
\usepackage{comment}
\usepackage{physics}
\usepackage{pdfpages}
\usepackage{xcolor}
\usepackage[english]{babel}
\makeatletter
\AtBeginDocument{\let\LS@rot\@undefined}
\makeatother
\usepackage[colorlinks=true]{hyperref}
\usepackage{graphicx}
\usepackage{sidecap}

\begin{document}


\newcommand{\TitleName}{Spectral Multiplexing of Rare-earth Emitters in a Co-doped Crystalline Membrane}
	\title{\TitleName}
	
	\newcommand{\AffGeneve}{Département de Physique Appliqu\'{e}e, Universit\'e de Gen\`{e}ve, CH-1211 Gen\`{e}ve, Switzerland}
	\newcommand{\AffNice}{Université c\^{o}te d’Azur, CNRS, Institut de Physique de Nice (INPHYNI), UMR 7010, Parc Valrose, Nice Cedex 2, France}

	\author{Alexander Ulanowski}
        \author{Johannes Fr\"uh}
        \author{Fabian Salamon}
        \author{Adrian Holz\"apfel}
        \author{Andreas Reiserer}
        \affiliation{
Max-Planck-Institute of Quantum Optics, Quantum Networks Group, Hans-Kopfermann-Stra{\ss}e 1, D-85748 Garching, Germany \\
And Technical University of Munich, TUM School of Natural Sciences and Munich Center for Quantum Science and Technology (MCQST), James-Franck-Stra{\ss}e 1, D-85748 Garching, Germany\\
Email Address: andreas.reiserer@tum.de
}

\date{\today}
	
\begin{abstract}
The spectral addressing of many individual rare-earth dopants in optical resonators offers great potential for realizing distributed quantum information processors. To this end, it is required to understand and control the spectral properties of the emitters in micron-scale devices. Here, erbium emitters are investigated in a Fabry-Perot resonator which contains a ten-micrometer-thin membrane of crystalline yttrium orthosilicate that is co-doped with europium. The co-doping allows for tailoring the inhomogeneous distribution of the emitter frequency. With this approach, we observe more than 360 spectrally resolved emitters with Purcell factors exceeding 35, each of which constitutes an individually addressable qubit within the micron-scale resonator. In addition to this spectral multiplexing, the optical coherence is preserved up to $\SI{0.62(3)}{\milli\second}$ under dynamical decoupling. Without decoupling, the coherence still reaches the lifetime limit for the emitters with the strongest Purcell enhancement that leads up to a 110-fold lifetime reduction, down to $\SI{0.104(9)}{\milli\second}$. Future work may combine this with long-lived nuclear spin memories, which makes the investigated co-doped membranes a promising platform for quantum repeaters and distributed quantum computers.
\end{abstract}

\maketitle 

\section{Introduction} \label{sec:Intro}

Rare-earth dopants are an emerging platform for distributed quantum information processing \cite{zhong_emerging_2019}. Pioneering experiments have used large ensembles of dopants to realize efficient quantum memories for single photons \cite{de_riedmatten_solid-state_2008}, to store entanglement in remote crystals \cite{clausen_quantum_2011, saglamyurek_broadband_2011}, and to implement long-term memories \cite{longdell_stopped_2005} with storage times exceeding six hours \cite{zhong_optically_2015}. Recently, however, also single dopants have been explored as emitters that enable efficient spin-photon interfaces for distributed quantum information processing systems. To this end, the long lifetimes of the optically excited states have been substantially reduced by integrating the emitters into optical resonators with large quality factor and small mode volume \cite{reiserer_colloquium_2022}. In this way, single dopants have been observed and controlled in several experimental systems \cite{dibos_atomic_2018, chen_parallel_2020, kindem_control_2020, ulanowski_spectral_2022, xia_tunable_2022, yu_frequency_2023, ourari_indistinguishable_2023, gritsch_purcell_2023, deshmukh_detection_2023}.

In all of these devices, the emitters are close to interfaces, with distances ranging between tens of nanometers and ten micrometers. It is therefore important to gain a detailed understanding of the spectral properties of rare-earth emitters in such a setting. In particular, one is interested in the characteristics of single emitters, such as their optical coherence and spectral diffusion, and in the shape of the inhomogeneous broadening. The latter is of critical relevance for realizing cavity-protected ensemble quantum memories \cite{putz_protecting_2014, zhong_interfacing_2017}, for understanding and controlling the coherence of spin transitions \cite{merkel_dynamical_2021}, and for spectral multiplexing \cite{chen_parallel_2020, ulanowski_spectral_2022}, i.e. the frequency-selective addressing of many emitters in the same resonator.

To achieve a high multiplexing capacity for single dopants, one needs a tailored spectral distribution: On the one hand, the spectral density should be large enough to have many emitters; on the other hand, it should not be too large - otherwise, the lines of individual emitters start to overlap such that they cannot be resolved spectrally. Implementing such a tailored spectral distribution is not straightforward: Even in nanoscale resonators, the frequency-multiplexed control of individual dopants requires precise doping on the sub-ppb level \cite{dibos_atomic_2018, chen_parallel_2020, kindem_control_2020, ulanowski_spectral_2022, xia_tunable_2022, yu_frequency_2023, ourari_indistinguishable_2023, gritsch_purcell_2023, deshmukh_detection_2023}. In a growth from the melt, this is hard to achieve -- in particular for the best-studied rare-earth host crystals, such as yttrium orthosilicate (YSO) \cite{dibos_atomic_2018, chen_parallel_2020, ulanowski_spectral_2022, horvath_extending_2019} and yttrium orthovanadate \cite{kindem_control_2020}. In these and other crystals containing rare-earth elements \cite{ahlefeldt_method_2013, berrington_negative_nodate}, trace impurities of the dopants typically exceed the desired concentration by orders of magnitude. Thus, previous experiments have worked in the far-out tails of the inhomogeneous distribution, where the spectral density is low. However, dopants in these sites will often be close to one or several unknown impurities that not only induce strain (which shifts the optical lines), but can also lead to uncontrolled decoherence and increased spectral diffusion. Furthermore, as the spectral density falls off quickly towards the tail of the distribution, only a limited number of emitters has been controlled with high fidelity \cite{chen_parallel_2020}. Possible workarounds are the use of nanocrystals \cite{deshmukh_detection_2023} or ultrapure crystals that can be grown without significant rare-earth impurities \cite{xia_tunable_2022, gritsch_purcell_2023, yu_frequency_2023, wang_single-electron_2023, ourari_indistinguishable_2023} and then doped by implantation to the desired concentration. In both cases, the choice of host is severely restricted, and the emitters are inevitably close to interfaces that deteriorate the coherence. 

In this work, we therefore investigate co-doping, i.e. adding another rare-earth species during the growth, as an alternative technique to achieve a tailored spectral density. To this end, we first perform a detailed characterization of the optical properties of individual rare-earth emitters in a ten-micrometer-thin crystalline membrane, bonded to a glass mirror that is part of a Fabry-Perot resonator with a high quality factor \cite{merkel_coherent_2020}. Combined with efficient single-photon detection, this facilitates a high signal-to-background ratio, over four orders of magnitude of the fluorescence signal. This allows for studying the inhomogeneous broadening of the optical transition frequency down to the level of single dopants. 

We then show that a co-doping concentration on the level of 100 ppm, a value that can be precisely controlled, allows for tailoring the inhomogeneous distribution to the desired spectral density by creating "satellite" side-peaks. Compared to the main line, the number of emitters in the satellites is reduced in proportion to the co-doping concentration. While previous works that investigated co-doping \cite{bottger_controlled_2008, thiel_optical_2012, welinski_effects_2017} used large dopant ensembles in macroscopic crystals, we investigate the spectral distribution on the level of single emitters in micrometer-sized membranes. Besides characterizing the statistics of their linewidths, we demonstrate that the adjusted spectral distribution allows for high-fidelity frequency-multiplexed control of individual emitters. At the same time, the optical coherence is preserved up to the lifetime limit in optical resonators at low magnetic field and a moderate temperature of $\lesssim \SI{2}{\kelvin}$. This relaxes the experimental requirements as compared to previous experiments \cite{kindem_control_2020, ulanowski_spectral_2022, ourari_indistinguishable_2023} and makes a larger class of hosts and emitters available for distributed quantum information processing.

In addition to tailoring the inhomogeneous distribution of the optical transition, our approach of co-doping offers further advantages. First, it is expected that it can extend the ground-state coherence of the dopants by broadening the inhomogeneous distribution of the spin transition, which can reduce the flip-flop rate \cite{car_optical_2019} and thus magnetic noise. Even more important, we find that the electronic spin transition frequency in the proximity of co-dopants is shifted relative to that of the isolated emitters by much more than the inhomogeneous linewidth. This reduces instantaneous spectral diffusion that limits the electronic spin coherence time even at very low doping concentration \cite{merkel_dynamical_2021}. Finally, co-doping gives access to advanced quantum information processing capabilities \cite{kinos_roadmap_2021}. In particular, conditional phase shifts between neighboring dopants  \cite{longdell_demonstration_2004}, also of different species, can be mediated by electrical and magnetic dipole interactions, or by strain \cite{louchet-chauvet_strain-mediated_2023}. This can enable quantum information processing in rare-earth-based devices \cite{kinos_roadmap_2021} and may give access to the storage of quantum states in memories with hour-long coherence times \cite{zhong_optically_2015}.

\section{Experimental setup}

We investigate erbium dopants, which are of special interest in the context of quantum networking as they are the only known emitters that combine second-long coherence of a ground state \cite{rancic_coherence_2018} with an optical transition at a telecommunication wavelength \cite{reiserer_colloquium_2022}. The latter can be extremely narrow \cite{bottger_optical_2006} and can even enable lifetime-limited coherence in optical resonators \cite{merkel_coherent_2020}. This is required for quantum interference of light that is a key enabler for quantum networking \cite{ruf_quantum_2021, reiserer_colloquium_2022}.

Because of these properties, single erbium dopants have been studied in a variety of setups: With nanophotonic resonators transferred to the surface of YSO \cite{dibos_atomic_2018, chen_parallel_2020} and other crystals \cite{yu_frequency_2023, ourari_indistinguishable_2023}, with nanophotonic resonators made from erbium-doped silicon \cite{gritsch_purcell_2023} or lithium niobate \cite{yang_controlling_2023}, as well as with yttrium oxide nanoparticles \cite{deshmukh_detection_2023} and thin  crystalline YSO membranes \cite{merkel_coherent_2020, ulanowski_spectral_2022} in high-finesse Fabry-Perot cavities. 

\begin{figure*}[tb]
\includegraphics[width=1\textwidth]{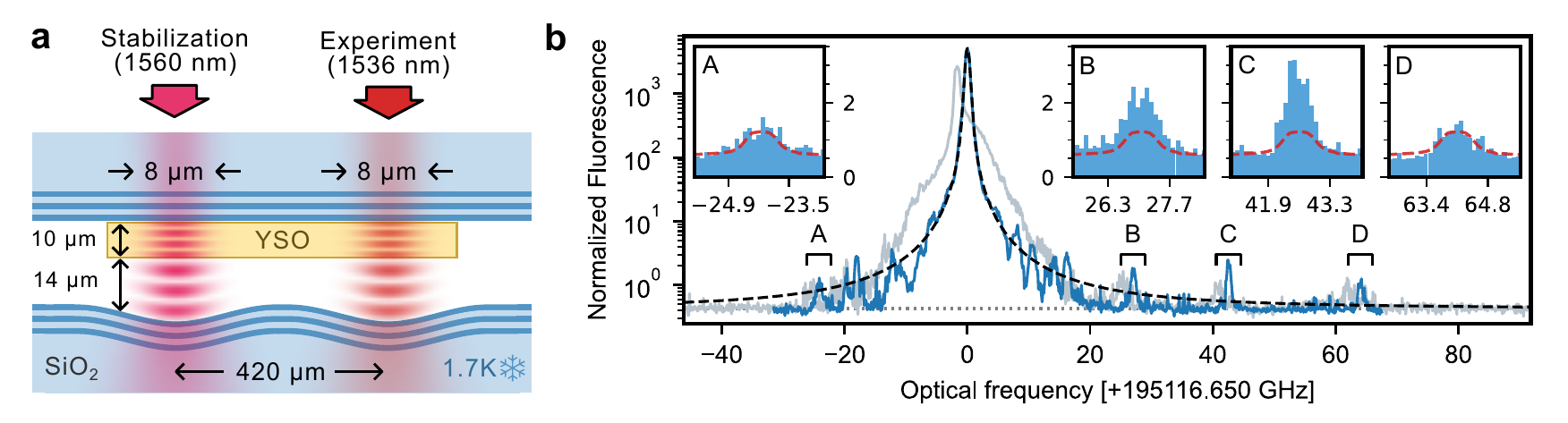}
\caption{\label{fig:Setup_inhomDist} 
a) Experimental setup (not to scale). A $\SI{10}{\micro\meter}$ thin membrane of YSO is integrated into an optical resonator with a mirror separation of $~\SI{24}{\micro\meter}$ and a mode waist of ${<\SI{4}{\micro\meter}}$. A second, close-by resonator allows for the stabilization of the mirror separation using a strong "stabilization" laser without significant heating of the "experiment" resonator that is kept at a variable temperature down to $\SI{1.7}{\kelvin}$ in a closed-cycle He cryostat. b) Fluorescence spectrum at zero magnetic field. The erbium ensemble exhibits an inhomogeneous broadening in the range of $\SI{1}{\giga\hertz}$. In the first cooldown after assembly, the line is broadened (gray), which is attributed to strain that originates from the thermal contraction. After several cooldowns, this strain relaxes (blue), giving a narrower linewidth that is well-fit by a Voigt profile (black, dashed) up to the regime in which the strain caused by the co-dopants is no longer continuous because of the discreteness of the crystal lattice \cite{stoneham_shapes_1969}. In this regime, sidepeaks are observed that originate from erbium emitters with a co-dopant in a close-by lattice site. The insets show a zoom into several of these side-peaks (blue bars) on a linear scale, and with a shifted copy of the central line (red, dashed). The latter is scaled by the co-dopant concentration, i.e. a factor of $10^{-4}$, before adding the independently measured average dark counts of the detectors.}
\end{figure*}

We follow the last approach, as it allows to keep the emitters further away from interfaces that can otherwise spoil their optical coherence. We thus investigate $\SI{10}{\micro\meter}$ thin membranes made from an yttrium-orthosilicate crystal by chemo-mechanical polishing as described previously \cite{merkel_coherent_2020}. The membrane is integrated in a Fabry-Perot resonator, as shown in Figure \ref{fig:Setup_inhomDist}a, to enhance the emission such that single-dopant detection and control becomes feasible \cite{ulanowski_spectral_2022}. Compared to this prior work, we improved the mirror parameters to increase the Purcell enhancement and enhance the outcoupling, as detailed in the Methods section. In addition, we now use a second resonator on the same mirror substrate for controlling the cavity length. This avoids that the stabilization laser heats the crystal at the location of the studied dopants.

From earlier measurements \cite{dibos_atomic_2018, ulanowski_spectral_2022}, it is known that undoped crystals of YSO contain trace impurities of erbium on the level of 0.1 to 1 ppm. In our setup, the resulting spectral density at the center of the inhomogeneous line is thus about four orders of magnitude too large for the control of individual dopants, as this requires that their average separation in frequency is significantly larger than their spectral diffusion linewidth of $< \SI{0.2}{\mega\hertz}$ found previously at large magnetic fields \cite{ulanowski_spectral_2022}. To adjust the spectral density to the requested level, we therefore start from crystals in which a fraction of $10^{-4}$ of the yttrium atoms is replaced by another species. We choose europium, another rare-earth dopant that features very long-lived nuclear spin states and no unpaired electronic spin. Thus, even at comparatively high doping concentration it is not expected to introduce excess magnetic noise to the crystal environment. Furthermore, its coherence time can exceed six hours~\cite{zhong_optically_2015}, making it a particular interesting candidate for coherent dopant-dopant-interactions and the implementation of a long-lived quantum network memory.

\section{Inhomogeneous distribution} \label{sec:BroadSpectrum}

To study the effects of the europium co-doping, we first determine the inhomogeneous linewidth of the erbium dopants in site 1 around $\SI{1536.48}{\nano\meter}$. To this end, we measure the fluorescence of the dopants after resonant excitation as a function of frequency. In this experiment, the frequency of the laser pulses and that of the resonator have to be swept synchronously. To implement this, the length of the cavity is stepwise adjusted by applying a voltage to a piezo tube that holds the mirrors and controls their distance. For each of the cavity length settings, all dopants whose optical transition frequency falls within the \SI{65}{\mega\hertz} FWHM linewidth of the resonator are excited by sweeping the frequency of the excitation laser pulses over a range of $\SI{100}{\mega\hertz}$ around the cavity resonance. Then, the fluorescence is recorded in a time interval of $\SI{0.25}{\milli\second}$ after the laser pulse has been switched off. This is repeated 800 times and averaged before changing the cavity length and center of the laser sweep to record the next data point. The resulting spectrum is shown in Figure \ref{fig:Setup_inhomDist}b. The fluorescence is normalized to display the number of resonant emitters, i.e. it is divided by the average number of photons detected from a single emitter in the experiments described below.

Because of the high signal-to-noise ratio enabled by superconducting nanowire single photon detectors (SNSPDs), we can observe the inhomogenous distribution over four orders of magnitude in fluorescence. In the first cooldown of the device, the observed distribution is asymmetric and broadened (gray), which we attribute to strain caused by the different thermal expansion coefficient of the membrane and the fused silica mirror to which it is bonded. However, after thermal cycling and simultaneously switching to a different resonator on the same mirror, the obtained distribution is more symmetric and does not change significantly in subsequent cooldowns of the device (blue).

If only strain caused by the size mismatch of the dopants and the host would cause the inhomogeneous broadening, one would expect a Lorentzian lineshape \cite{stoneham_shapes_1969}. We find that this only matches the curve at large detunings. At the center of the distribution, however, a Lorentzian fit significantly underestimates its width. This indicates that several effects contribute to the observed broadening, some of which exhibit a steeper, e.g. Gaussian, decay. We thus find a good agreement when fitting a Voigt profile to the data - a convolution of a Lorentzian (with $\gamma=\SI{0.14(1)}{\giga\hertz}$ FWHM) and a Gaussian (with $\sigma=\SI{0.27(1)}{\giga\hertz}$ FWHM). The Lorentzian contribution is in good agreement with the expectation  from earlier measurements of Er:Eu:YSO that found a Lorentzian broadening with a linear concentration dependence of $\SI{11}{\giga\hertz}$ per percent of europium \cite{bottger_controlled_2008}. This would correspond to a broadening of $\SI{0.11}{\giga\hertz}$ for the 100 ppm europium concentration used in this work. 

The additional Gaussian contribution is required to achieve a good fit to the data at small detunings. Its FWHM is comparable to that obtained in nominally undoped membranes \cite{merkel_coherent_2020} and in weakly-doped bulk crystals with dimensions of several millimeters \cite{bottger_optical_2006, lauritzen_state_2008, thiel_rare-earth-doped_2011, cova_farina_coherent_2021}. Thus, our measurements further illustrate that the inhomogeneous broadening in Czochralski-grown YSO crystals originates from fluctuations on microscopic rather than macroscopic length scales.

The fast drop of the inhomogeneous distribution in its tails would limit the number of isolated, single emitters that can be controlled. This is overcome by the co-doping, as replica of the central line with a reduced spectral density are obtained on both sides of the distribution. These "satellite" lines emerge as the co-dopants can only be integrated at discrete lattice sites. This results in a discrete nature of the generated strain at surrounding erbium sites. As the strain will fall off with the distance between the dopant and co-dopant, satellite peaks are only resolved if the two are in close proximity.

As europium atoms can be integrated in many locations surrounding a given erbium dopant while creating different strain fields, there is a large number of satellite lines. The outermost ones are best-resolved as they are clearly separated from the main line. A zoom into these is shown in the insets of Figure \ref{fig:Setup_inhomDist}b (blue data). For comparison, we also show a copy of the main line (red), shifted in frequency and rescaled by the europium concentration of $10^{-4}$ before adding the independently measured average dark counts of the detector. The width of the satellite peaks resembles the center line very well. For two of them, inset A and D, the amplitude fully matches the expectation, such that they can both be attributed to europium dopants on one well-defined site neighboring the erbium emitters. Precisely determining this site is hindered by the expected anisotropy of the generated strain field, but could be achieved with additional measurements of e.g. the dipolar interaction strength \cite{ahlefeldt_method_2013}. The other observed satellite peaks are multiples of the outermost ones, with approximately two- or three times the fluorescence, which may indicate that several locations of the co-dopants lead to a similar shift of the erbium emission.

The quantitative agreement of the satellite line amplitude with the expectation suggests that erbium and europium are randomly integrated into the lattice, and that there is no strong preference for the two to form clusters in the used Czochralski growth process. Thus, by adjusting the concentration of Eu we can obtain a tailored spectral density in the sidepeaks that  -- in the current sample -- reaches four orders of magnitude lower than in nominally undoped crystals.

\section{Properties of single emitters} \label{sec:single emitter properties}

\begin{figure*}[tb!]
\includegraphics[width=1\textwidth]{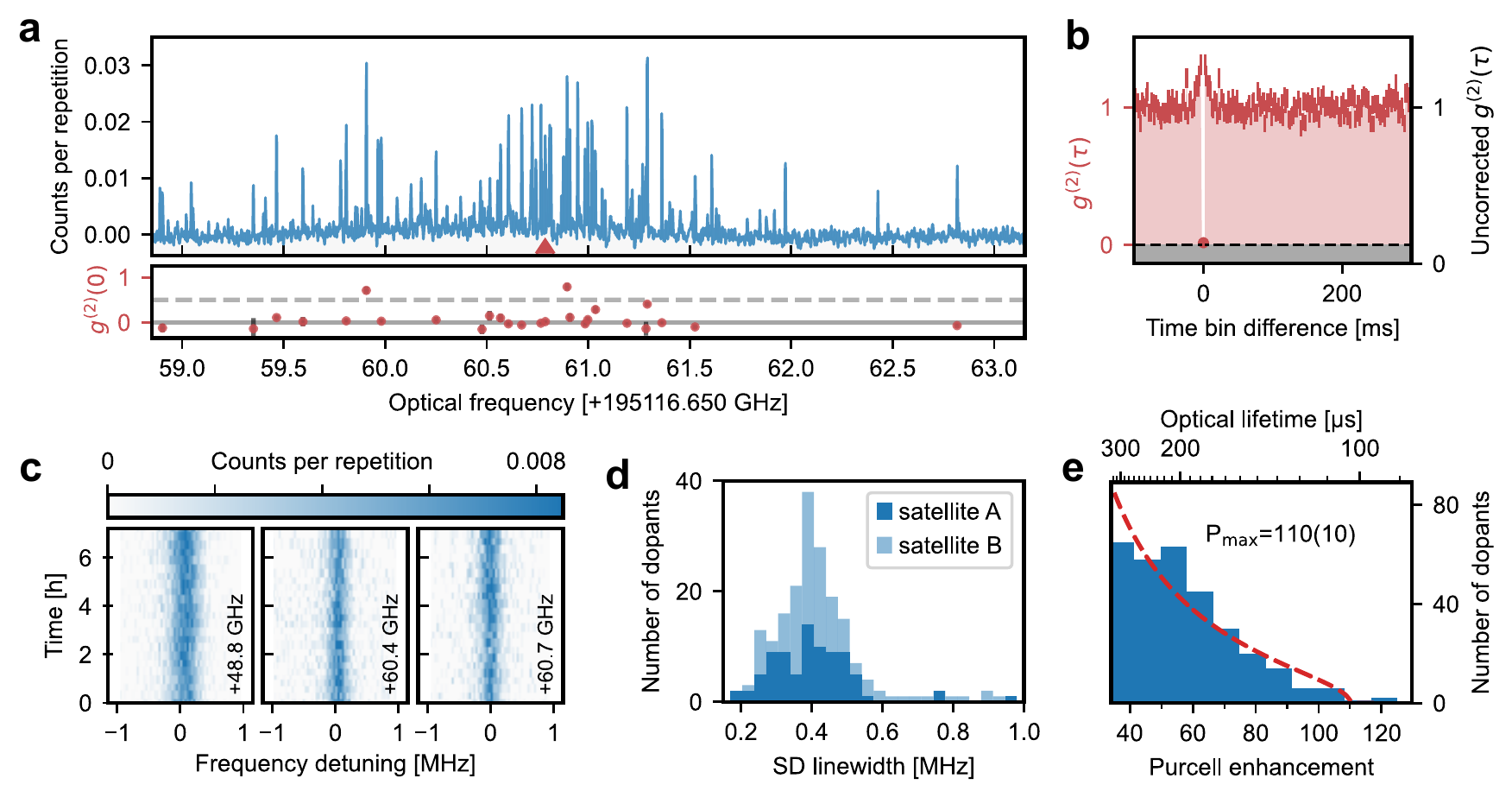}
\caption{\label{fig:SpectralProperties_Singles} 
Spectral addressing of individual emitters in the satellite lines. a) Top panel: High-resolution fluorescence spectroscopy of the outermost satellite line "D", in which about 50 single dopants exhibit a strong Purcell enhancement. Thus, their emission is channeled into the resonator, leading to the detection of single photons after the excitation pulse with up to $~3\%$ probability (peaks). Bottom panel: On most of the studied peaks, the light emitted after pulsed excitation exhibits very low values -- close to zero (gray line) -- of the dark-count corrected autocorrelation function at zero delay $g^{(2)}(0)$. This proves that many single emitters, ${g^{(2)}(0)<1/2}$ (gray dashed line), can be spectrally resolved. b) Autocorrelation measurement on a randomly chosen dopant (red triangle in a). The finite value at zero delay is explained by the independently measured dark counts of the detector (gray area, black dashed line). Subtracting this contribution (red left axis) leads to almost perfect antibunching at zero delay. c) Spectral stability of three randomly chosen emitters. The fluorescence after excitation with narrowband Gaussian pulses exhibits no blinking, bleaching or significant spectral shifts over many hours. d) Distribution of the spectral diffusion (SD) linewidth at a low magnetic field of approximately $\SI{0.2}{\milli\tesla}$. The fitted linewidth of each dopant depends on the magnetic field fluctuations within the host crystal. These are of similar size, but differ slightly between dopants, leading to different SD linewidth. As no significant difference is observed between emitters in different satellite peaks, the data of peak A and B is summed in the same histogram (dark and light blue, respectively). e) The distribution of the lifetime (top axis) and Purcell enhancement factors (bottom axis) of the dopants with the strongest coupling to the resonator is in good agreement with the theoretical expectation (red dashed). A total of $~360$ emitters with an enhancement $> 35$ can be resolved spectrally in the satellite peaks B-D.
}
\end{figure*}

In the following, we study the properties of single erbium emitters in the satellite lines. To this end, we perform a high-resolution measurement of the fluorescence as a function of the laser frequency. As individual emitters are expected to be very narrow, both the resonator and the excitation laser are stabilized in this experiment using a frequency comb as a reference \cite{ulanowski_spectral_2022}. To efficiently excite the emitters we use the Rapid Adiabatic Passage technique by applying frequency-chirped square pulses of \SI{4}{\micro\second} duration with a chirp of \SI{0.25}{\mega\hertz\per\micro\second}. A magnetic field of $\SI{0.35}{\tesla}$ is applied approximately along the b-axis of the crystal {, where the effective g-factors of ground and excited states are $g_g\simeq 9$ and $g_e\simeq 10$, respectively \cite{sun_magnetic_2008}}. The b-axis is chosen as the Zeeman transitions of both magnetic classes are split by roughly the same amount, $\approx\SI{6}{\giga\hertz}$. Thus, only one spin level is excited (as discussed below). Within a span of $\SI{3}{\giga\hertz}$, we find about fifty peaks in the satellite line "D", as shown in the upper panel of \ref{fig:SpectralProperties_Singles}a. 

To demonstrate that these peaks correspond to the emission of individual dopants, we measure the photon statistics when exciting a randomly-chosen emitter with chirped pulses in which the laser is swept over $\SI{2.5}{\mega\hertz}$ with a chirp rate of \SI{0.6}{\mega\hertz\per\micro\second}. As the dead time of the used detector, $\lesssim \SI{0.1}{\micro\second}$, is much shorter than the Purcell-enhanced lifetime of the emitters, $\gtrsim \SI{0.1}{\milli\second}$, a single detector is sufficient to measure the autocorrelation function $g^{(2)}(\tau)$. We observe almost perfect antibunching, see Figure \ref{fig:SpectralProperties_Singles}b. The finite value at zero delay, $g^{(2)}(0)=0.13(2)$, is consistent with the contribution of accidental coincidences that originate from detector dark counts \cite{becher_nonclassical_2001} and could be reduced by two orders of magnitude with better nanowire devices \cite{akhlaghi_waveguide_2015}. The small bunching shoulders are attributed to spin pumping, as studied in detail below.

In the lower panel of Figure \ref{fig:SpectralProperties_Singles}a, we show the result of the autocorrelation measurements of all emitters in the satellite peak "D" that exhibit a strong Purcell enhancement $P>30$ and thus a large number of detected photons, enabling a measurement of the autocorrelation function with a standard deviation below a threshold of 0.3 within 7 minutes. We then subtract the dark count contribution \cite{becher_nonclassical_2001} to investigate how well-isolated the emitters are. We find that almost all the dopants exhibit very low values of the corrected $g^{(2)}(0)<0.15$. This demonstrates that the co-doping introduced in this work allows for tailoring the spectral density such that the background of weakly coupled dopants observed in our earlier measurements \cite{ulanowski_spectral_2022} is negligible, while still enabling spectrally multiplexed addressing of many emitters with strong Purcell enhancement. 

This multiplexing requires that the average frequency separation of the peaks is much larger than their average width. To determine the latter with high resolution, we repeatedly sweep the laser over the same range and investigate the fluorescence signal in the first $\SI{0.25}{\milli\second}$ after the excitation laser pulse, averaged over 3000 repetitions per data point. We thus obtain a full spectrum every 20 minutes. As shown in Figure \ref{fig:SpectralProperties_Singles}c, we find that the erbium emitters are very stable over many hours. They do not exhibit any blinking or bleaching. However, the width of the spectral features is much larger than the lifetime limit that is below $\lesssim \SI{1}{\kilo\hertz}$. This indicates that on timescales shorter than that of the measurement, the dopants exhibit significant spectral diffusion caused by electric or magnetic field fluctuations. In our earlier work on nominally undoped membranes, we found that the latter are dominant. They have two contributions: The nuclear spin bath, predominantly caused by the yttrium nuclei surrounding each erbium dopant with 100\% isotopic abundance, as well as paramagnetic impurities. When the latter are frozen in a large magnetic field, linewidths down to $\SI{0.15}{\mega\hertz}$ have been observed \cite{ulanowski_spectral_2022}. Here, we instead investigate the regime of low magnetic field, such that the effect of paramagnetic impurities can be studied. By fitting the signal of the dopants, averaged over 10 minutes, to a Gaussian function, we obtain the spectral diffusion (SD) linewidth. Its FWHM is in the range of 0.2 to $\SI{1}{\mega\hertz}$ for all investigated dopants. A histogram of the SD linewidth distribution is shown in Figure \ref{fig:SpectralProperties_Singles}d. The observation that the SD differs between the emitters demonstrates that the noise is local to each dopant. As the yttrium nuclear spin bath is identical for each emitter, and was found to give a lower contribution in our earlier work, we conclude that the SD at low magnetic field is dominated by paramagnetic impurities. Still, their effect is quite moderate, such that the measured SD linewidth is among the lowest reported in the literature, comparable to that in \cite{ulanowski_spectral_2022, ourari_indistinguishable_2023}.

Besides the SD linewidth, also the distribution of the Purcell enhancement factors deserves further study. To this end, the decay of the single-dopant fluorescence is measured after pulsed excitation for all emitters that are clearly resolved in the satellite lines B-D. Similar results would be expected for measurements on the other satellite lines. A histogram of the lifetimes, determined from exponential fits, is shown in Figure \ref{fig:SpectralProperties_Singles}e. The measured Purcell factors match the distribution calculated from the electric field amplitude of the fundamental Gaussian standing-wave mode of the Fabry-Perot cavity (red dashed) \cite{merkel_coherent_2020}, with a fitted density of emitters in the satellite line and a maximum Purcell enhancement of 110(10).

\section{Spin properties}
\begin{figure*}[tb!]
\includegraphics[width=1\textwidth]{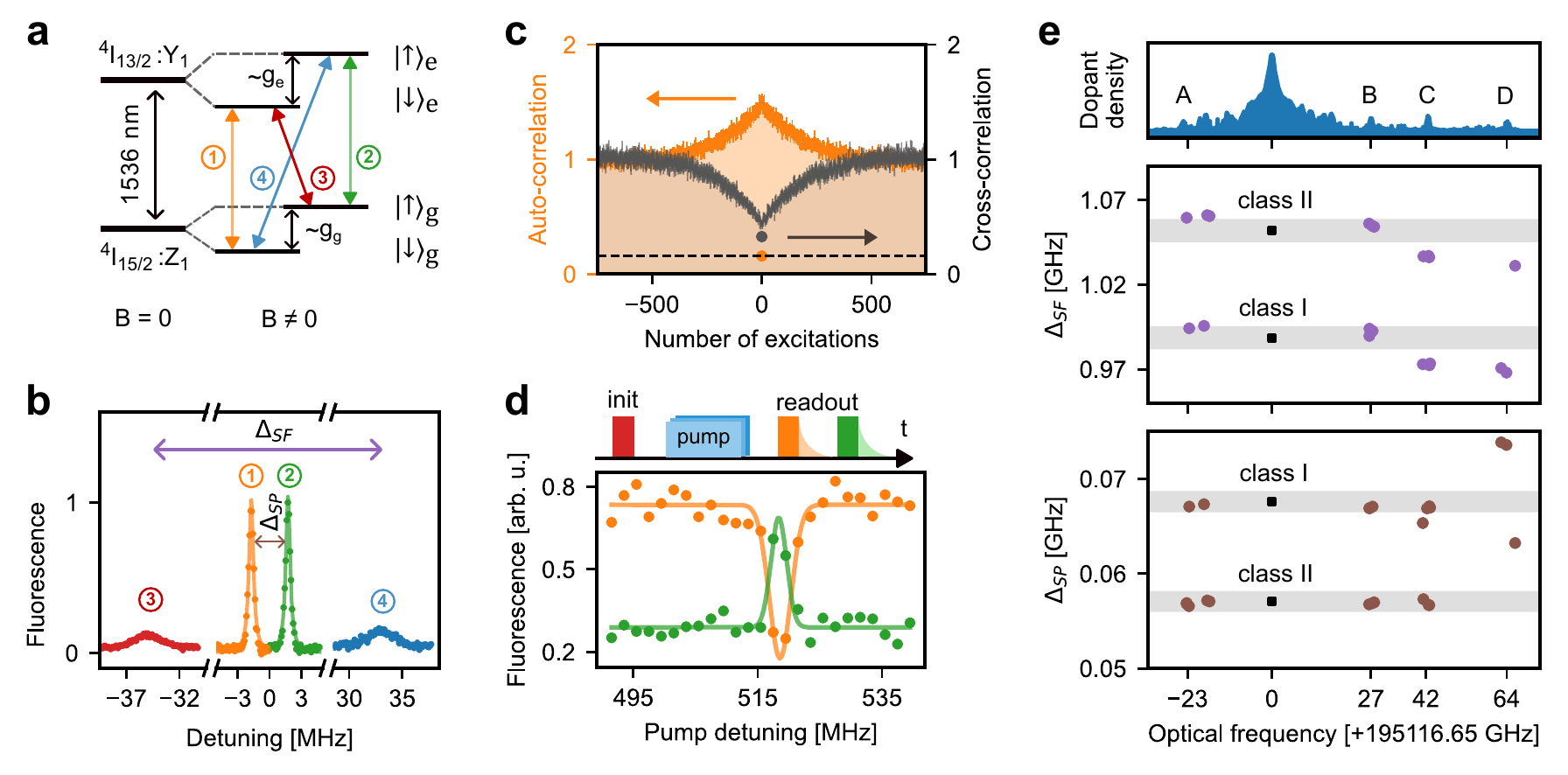}
\caption{\label{fig:SpinProperties}
Spin properties of individual emitters. a) Level scheme. The lowest crystal field level $Z_1$ of the $^4I_{15/2}$ ground state, and the lowest crystal field level $Y_1$ of the optically excited $^4I_{13/2}$ state split in an applied magnetic field, yielding four transitions (colored arrows) at different frequency. b) When a magnetic field is applied, the two spin-preserving (green, orange) and spin-flip (red, blue) lines can be resolved. The latter are broader, as the corresponding transitions are more sensitive to magnetic field noise from the nuclear spin bath. c) Photon correlation functions in a small magnetic field. Laser pulses selectively excite the spin-preserving transition of a single dopant, leading to the emission of antibunched light at zero delay, as discussed in detail in Fig. \ref{fig:SpectralProperties_Singles}b. The main difference shown in this panel is an additional, pronounced bunching feature observed at short time difference when only the emission after pulses of the same frequency is analyzed (orange). In contrast, when looking at the cross-correlation, i.e. light emitted after pulses of alternating frequency, a broad antibunching feature is observed. This proves the spin-selective excitation and the spin-preserving character of the transitions. d) Top: Pulse sequence to determine the splittings of the transition frequencies. Bottom: Measurement to determine the frequency of the blue (4) spin-flip transition of a single dopant in satellite line "C". e) The splittings of the spin-preserving ($\triangle_\text{SP}$, bottom) and spin-flip ($\triangle_\text{SF}$, center) transitions depend on the detuning from the center of the inhomogeneous broadening. The difference between the satellite lines (top) is larger than the FWHM of the spin inhomogeneous broadening, which is measured on the central line using a different technique, persistent spectral holeburning (grey shaded areas).}
\end{figure*}

To study the spin properties of the emitters, we again apply a small magnetic bias field along the $b$ direction of the YSO crystal. This splits the effective spin levels of the $^4I_{15/2}$ ground and the $^4I_{13/2}$ optically excited state, as shown in Figure \ref{fig:SpinProperties}a. We first apply a field of $\SI{0.2}{\milli\tesla}$ and measure the fluorescence of a single dopant as a function of the laser detuning. In Figure \ref{fig:SpinProperties}b, one can clearly resolve the spin-flip (red and blue) and spin-preserving (orange and green) transitions. The former are less intense owing to the different magnitude of the electric dipole transition rate \cite{liu_spectroscopic_2005} that leads to a lower cyclicity of the transition \cite{raha_optical_2020}. In addition, the SF transitions are broader, $\SI{3.2(3)}{\mega\hertz}$ as compared to $\SI{0.50(2)}{\mega\hertz}$ for the SP ones. This strengthens our hypothesis that magnetic field fluctuations caused by electronic and nuclear spins are the dominant source of SD, as the spin-flip transitions are more sensitive to them. A quantitative comparison between the difference in the splitting and the difference in the linewidth, however, is not appropriate, as the full interaction Hamiltonian needs to be evaluated to predict the broadening in the regime of moderate B field, in which the interaction terms dominate in the Hamiltonian \cite{car_selective_2018, merkel_dynamical_2021}.

To further study the spin dynamics upon optical excitation, we apply a field of $\SI{1}{\milli\tesla}$. We then excite a randomly chosen dopant with laser pulses whose bandwidth of $\SI{1}{\mega\hertz}$ is chosen such that the emitter is efficiently excited on one of the transitions, but not excited on the other. We then analyze the statistics of the light emitted within a window of $\SI{400}{\micro\second}$ after the pulses, See Figure \ref{fig:SpinProperties}c. We alternate between pulses on each of the two spin-preserving (SP) transitions, orange and green in Figure \ref{fig:SpinProperties}a and b. When only evaluating the counts after the pulses exciting one of the transitions, e.g. the one between the lower spin levels (orange), we again observe perfect antibunching in the auto-correlation function, as discussed earlier. However, in addition to the antibunching feature at zero delay, we also observe broader bunching shoulders. The reason is that only one of the spin states is excited ("bright"), while the other is off-resonant and thus "dark". Detection of an emitted photon signals that the spin is in the bright state, which leads to a higher detection probability after subsequent excitation pulses at the same frequency. At longer delay, the spin has decayed to a random state. As only one of the transitions is resonant with the laser, the average number of detected photons is thus reduced. In effect, this results in "bunching" shoulders in the $g^{(2)}$ function.

The bunching is more pronounced than that measured at larger magnetic fields, see e.g. Fig. \ref{fig:SpectralProperties_Singles}b, where it is limited by a short spin lifetime \cite{bottger_optical_2006}. At low fields, instead, the spin state is randomized with increasing number of excitations because the emitter can also decay on the spin-flip (SF) transition, even though the latter is not enhanced by the Purcell effect because of its large detuning with respect to the cavity resonance. In the depicted measurement, the resulting decay of the bunching is exponential in the number of repetitions, with a decay constant of 205(5) attempts. This number could be further increased by aligning the optical dipole to the cavity mode by applying the magnetic field with an optimized orientation \cite{raha_optical_2020}. This would enable single-shot readout of the spin state with a high fidelity.

In addition to the autocorrelation, one can also evaluate the cross-correlation when alternatingly exciting the SP transitions. This then corresponds to the probability to detect a photon after driving one SP transition, conditioned on a detection event after a prior pulse driving the other SP line. As expected, the resulting curve (gray in Figure \ref{fig:SpinProperties}c) exhibits an antibunching feature that decays inversely to the bunching feature of the autocorrelation measurement (orange).

With the ability to resolve the spin splitting, we can further determine if it differs between the satellite peaks. Possible reasons for such a discrepancy are that the Eu atoms may change the magnetic environment in their surrounding and that they distort the erbium site, leading to a change in the crystal field levels and thus the effective g-factors. To investigate this, we apply a slightly larger field, around $\SI{2.5}{\milli\tesla}$, and measure the splitting of the lines for several dopants. As the SF transitions at this field are far-detuned from the cavity resonance -- more than ten linewidths -- and thus experience no Purcell enhancement, we use a different scheme to determine their relative frequency, shown in Figure \ref{fig:SpinProperties}d. We first initialize ("init") the spin by driving the red SF transition in a broad spectral range with frequency-modulated laser pulses. Then, the blue SF transition is driven with a narrowband pulse ("pump"). Fitting the result of subsequent measurements on the SP transitions ("readout") allows us to determine the frequency of the SF transition with high accuracy.

In the analysis of the data, we observe two classes of emitters with slightly larger or smaller splitting, see Figure \ref{fig:SpinProperties}e. This is caused by a slight misalignment of the magnetic field from the $b$-axis, which entails that the two magnetic subclasses of erbium in YSO (commonly labeled I and II) are not fully degenerate. However, in addition to the difference between the classes, we observe that the splitting of the transitions differs between the satellite peaks, in particular for the outermost ones. The absolute difference is more pronounced on the SF transition (that is more sensitive to the B-field), while the relative change is stronger on the SP transition.

For comparison, we also show the splitting of persistent spectral holes burnt in the center of the inhomogeneous distribution, where many emitters contribute to the signal. For the outermost satellite peak "D", the difference of the spin-flip transition of single dopants is significantly larger than the FWHM of the holes in the central line (shaded areas) that is determined by the inhomogeneous broadening of the spin transitions. This has two favorable consequences: First, we expect that the spin lifetime will be increased in the satellite lines, as its dominant limitation at cryogenic temperature is the flip-flops with other spins \cite{car_optical_2019}. Those will be suppressed for the dopants in the direct vicinity of a co-dopant, as they are off-resonant with most other erbium emitters in the crystal. Second, the difference in the SF transition frequency entails that dopants in the satellite lines can be selectively addressed by microwave radiation. Thus, one can dramatically reduce the main limitation of the coherence of the dopants, which is instantaneous spectral diffusion that cannot be overcome by dynamical decoupling for dopants with anisotropic interactions \cite{merkel_dynamical_2021}. As the instantaneous-diffusion linewidth depends linearly on the dopant concentration, we expect that an improvement by four orders of magnitude is possible at the studied concentration of co-dopants using appropriate decoupling sequences \cite{merkel_dynamical_2021}. Considering that a coherence time exceeding $\SI{0.1}{\milli\second}$ has been observed for erbium in nominally undoped YSO \cite{chen_parallel_2020}, second-long coherence seems thus achievable on the electronic spin transition in the satellite lines.

\section{Optical coherence}

\begin{figure*}[tb]
\includegraphics[width=1\textwidth]{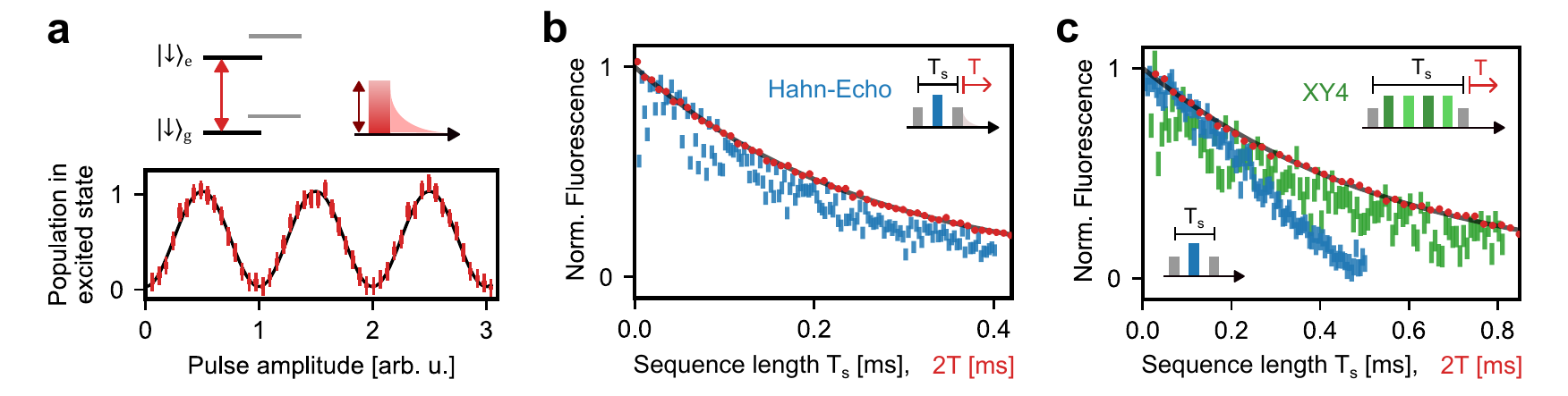}
\caption{\label{fig:OpticalCoherence} 
Optical coherence of emitters in the satellite lines. a) When the amplitude of the excitation pulses is swept, the fluorescence after the \SI{0.32}{\micro\second} long excitation pulse, which is proportional to the population in the optically excited state, exhibits Rabi oscillations without any signatures of damping. b) Coherence measured in a Hahn echo sequence (inset) on the optical transition. We first study an emitter with a short Purcell-enhanced decay of $\SI{131(1)}{\micro\second}$ (red fluorescence decay data and black fit curve). Here, two times the lifetime sets an upper bound to the envelope of the Hahn echo decay (blue) as a function of the sequence length. Thus, the studied emitter exhibits lifetime-limited coherence. c) The Hahn echo (blue) of a dopant with a longer lifetime of $\SI{290(10)}{\micro\second}$ (red) exhibits a steeper decay. Application of four optical pulses with alternating phase along x and y (inset) extends the optical coherence (green) back to the lifetime limit (red).}
\end{figure*}

After investigating the spin properties of the emitters, we now turn to their optical coherence. To this end, we further increase the magnetic field to $>\SI{250}{\milli\tesla}$ to avoid superhyperfine-induced decoherence, which happens on a timescale of seven microseconds at zero field \cite{car_superhyperfine_2020}. We then select a random dopant with a large Purcell-factor and apply Gaussian pulses with a FWHM of $\SI{0.32}{\micro\second}$ on the lower SP transition, as shown in Figure \ref{fig:OpticalCoherence} (top). When varying the pulse amplitude, we obtain Rabi oscillations without significant decay (bottom). This testifies that the coherence time is much longer than the pulse duration, and that the cavity resonance frequency and thus the laser power in the resonator is well stabilized. The result of such a measurement is very similar for all investigated dopants, except for the Rabi frequency that depends on the Purcell factor.

To determine the optical coherence time, we apply a Hahn echo sequence on the optical transition, see the inset of Figure \ref{fig:OpticalCoherence}b. It starts with an optical $\pi/2$ pulse. A subsequent $\pi$ pulse compensates for a static detuning of the emitter before another $\pi/2$ pulse is applied. We then evaluate the fluorescence after the last pulse as a function of the delay between the pulses. To exclude artifacts from pulse imperfections, we plot the difference between a measurement with unchanged phase and a measurement with inverted phase of the first pulse \cite{ulanowski_spectral_2022}. We find that for some of the dopants with large Purcell enhancement, the optical coherence $T_2$ exhibits a decay (main graph, blue data of a randomly-selected emitter) whose timescale is only bounded by the Purcell-enhanced lifetime $T_1=\SI{131(1)}{\micro\second}$ (red data and grey fit curve), realizing the equality sign in the relation $T_2 \le 2 \cdot T_1$. This observation of lifetime-limited optical coherence is a critical capability for the entanglement of dopants via the interference of photons \cite{reiserer_colloquium_2022}. The small remaining envelope modulation is caused by superhyperfine interactions. It can be eliminated by further increasing the magnetic bias field \cite{car_superhyperfine_2020}.

Because of the long lifetime of europium nuclear spins\cite{zhong_optically_2015}, no difference of the optical coherence is expected between the main line and any of the sidepeaks. However, we observe that it differs between individual dopants, which means that the noise that leads to dephasing originates in the immediate surrounding of each emitter. Figure \ref{fig:OpticalCoherence}c shows a measurement on another, randomly chosen emitter with a longer lifetime (red). In the Hahn echo measurement (blue), this emitter does not exhibit lifetime-limited coherence. However, the coherence can be extended by applying a pulse sequence on the optical transition that decouples quasistatic fluctuations caused by other impurities. Using XY4, a sequence of four $\pi$ pulses with a phase that alternates by $\pi/2$, the coherence is again extended to the lifetime limit of $T_2=\SI{0.62(3)}{\milli\second}$. This constitutes a fivefold increase compared to our previous work that achieved $\SI{0.12}{\milli\second}$ at large magnetic fields that were applied to freeze out spin noise \cite{ulanowski_spectral_2022}. The observed improvement is attributed to the new resonator stabilization scheme that avoids heating of the crystal at the position of the dopants.

\section{Conclusion} \label{sec:discussion}

In summary, we have demonstrated that co-doping is an effective technique to tailor the spectral density of rare-earth emitters in a common host material. At the studied concentration, it does not broaden the inhomogeneous distribution compared to that measured in bulk crystals \cite{bottger_optical_2006, lauritzen_state_2008, thiel_rare-earth-doped_2011, cova_farina_coherent_2021} and thicker membranes \cite{merkel_coherent_2020}. Still, in the satellite lines, it allows to resolve and control more than 360 individual emitters with Purcell enhancement factors $> 35$ in a ten micrometer thin membrane. These emitters can exhibit lifetime-limited optical coherence, and a spectral diffusion below $\SI{1}{\mega\hertz}$ that is dominated by magnetic field noise of the surrounding spins in the crystal. As the spin transitions of the satellite peaks are shifted with respect to those in the main line, we expect that second-long coherence may be achieved by dynamical decoupling. 

These advantages are not expected for emitters in the tail of the inhomogeneously broadened linewidth studied in previous experiments \cite{raha_optical_2020,  chen_parallel_2020, ulanowski_spectral_2022}. Such emitters can have interactions with one or several random impurity atoms in the crystal \cite{kornher_sensing_2020, uysal_coherent_2023}. In contrast to these experiments and others that use nuclear spin registers \cite{ruskuc_nuclear_2022}, our approach enables a reproducible and controlled coupling to one specific co-dopant at a predetermined position. When using elements with a nuclear spin, such as Eu, this paves the way to very long-lived quantum memories in the proximity of individual telecom emitters. This could be implemented based on the magnetic dipole interaction and the control techniques pioneered with nitrogen-vacancy centers in diamond \cite{taminiau_universal_2014}. However, in case magnetic dipolar couplings are used, the always-on character of the interaction Hamiltonian would limit the performance of quantum network memories \cite{reiserer_robust_2016}. This can be avoided in our approach by operating the system at a bias field that decouples the magnetic interaction \cite{zhong_optically_2015}, and instead using strain- \cite{louchet-chauvet_strain-mediated_2023} or electric-dipole coupling of the emitters in the optically excited state \cite{longdell_demonstration_2004, kinos_roadmap_2021}. To this end, it will be advantageous if both the dopant and the co-dopant allow for single-emitter addressing. In the current setup, this should be feasible by applying another control laser at $\SI{580}{\nano\meter}$ that is resonant with the $^7F_0$ to $^5D_0$ transition of europium. Based on this capacity, combining the excellent memory performance of Eu with the telecom transition of erbium emitters constitutes a promising system for the implementation of fiber-based quantum networks on a global scale.

\section{Methods} \label{sec:methods}

The experimental setup is similar to that used in our earlier work \cite{merkel_coherent_2020}. The supplementary information of this article contains a detailed description of the fabrication and mounting process of the thin crystal membrane and Fabry-Perot cavity. In the following, we will focus on the technical improvements and modifications of the setup that have been performed for the experiments described above. Besides the use of a co-doped crystal, this includes improvements to the Fabry-Perot cavity and its stabilization.

\subsection{Yttrium-orthosilicate sample}
Yttrium orthosilicate is generally considered a very favorable host for rare earth dopants for quantum applications for several reasons: First, as all rare earth dopants feature similar size and chemical properties, they can easily integrate into the crystal by replacing a single yttrium ion without causing major distortions of the crystal structure. Second, the low density of magnetic moments in YSO and the low crystal symmetry enable exceptional coherence properties, with the longest coherence times reported in any material for qubits encoded both in spin degrees of freedom \cite{zhong_optically_2015} and optical excitations \cite{bottger_optical_2006}.

However, as it is difficult to avoid trace impurities of other rare-earth elements in the starting yttrium material, YSO growth with a doping concentration at the required level of sub-ppb has not been demonstrated. Instead, the 100-ppm Eu-doped crystal used in this work has been grown by Scientific Materials using the Czochralski method (S/N 2-427-17). Crystals of this supplier are known to be very pure, with an erbium concentration below $1~\mathrm{ppm}$ \cite{dibos_atomic_2018, ulanowski_spectral_2022}.

To fabricate the membranes, crystals with dimensions of $\SI{5}{\milli\meter}\times\SI{5.5}{\milli\meter}$ are polished to a thickness of $10\pm1~\mathrm{\mu m}$ and a surface roughness below \SI{0.3}{\nano\meter} (rms) using a custom chemo-mechanical procedure as outlined in the supplementary material of \cite{merkel_coherent_2020}. Integrating it into a resonator of high-finesse, described below, allows for significant Purcell enhancement while avoiding detrimental effects of surface charges that spoil the spectral properties and coherence of the optical transitions in nanocavity-coupled YSO \cite{chen_parallel_2020}.

\subsection{Fabry-Perot cavity}
The cavity is formed by two fused-silicon mirror substrates with a diameter $d=\SI{7}{\milli\meter}$, super-polished to a roughness below \SI{0.2}{\nano\meter} (rms) by Research Electro Optics. One of the mirrors is flat, while a regular array of smooth concave depressions is fabricated on the other via laser ablation \cite{merkel_coherent_2020}. Each of these dimples forms a cavity that exhibits a Gaussian fundamental mode with its waist at the flat mirror surface. Here, the YSO sample is attached by van-der-Waals forces. 

To achieve the required reflectivity, both mirror substrates are coated prior to assembly with alternating layers of Nb$_2$O$_5$ and SiO$_2$ (Laseroptik GmbH). The dimpled mirror has a total of 31 layers and the flat outcoupling mirror has 27 layers. Compared to our previous experiments \cite{merkel_coherent_2020, ulanowski_spectral_2022}, the reduced reflectivity of the second mirror leads to an increased outcoupling efficiency of $75(5)\%$ while preserving a narrow cavity linewidth of $\Delta\nu=\SI{65(10)}{\mega\hertz}$ FWHM.

To still achieve a stronger Purcell enhancement, the radius of curvature of the dimples, $R=\SI{65(2)}{\micro\meter}$, is chosen more than a factor of two smaller than before. Together with a twofold reduction of the cavity length (to a mirror distance of $\SI{24}{\micro\meter}$), this allows for a stronger mode confinement, with a waist of $\SI{3.82(4)}{\micro\meter}$. For an emitter at the center of the cavity mode this results in a lifetime reduction up to a factor of $P=116$ when the light is polarized along the optical D2 axis of the crystal. In the experiment, Purcell factors up to $P=110(10)$ are observed, which constitutes an almost two-fold increase compared to our previous experiments \cite{ulanowski_spectral_2022}.

\subsection{Resonator stabilization}
During the experiment, the cavity needs to be stabilized using a piezo tube to the transition of the emitters. Previously, this was done using the Pound-Drever-Hall technique with a second laser on a higher resonator mode, which is separated by 28 nm from the erbium resonance in the current setup. However, the induced heating then reduces the optical coherence \cite{merkel_coherent_2020}. Thus, in this work instead of a higher mode of the same cavity, we use a second, close-by resonator. As both are located on the same mirror substrate, stabilizing one resonator also stabilizes the length of the other. However, the heating is spatially separated, such that the dopants can be probed at the base temperature of the used closed-cycle cryostat (AttoDry 2100). 

Apart from the changed geometrical arrangement, the used stabilization setup is the same as before. However, adding additional vibration dampers (Newport VIB320) to the outer vacuum chamber, where the resonator is mounted, leads to an improved mechanical stability, with residual fluctuations of $\SI{1.5}{\pico\meter}$ rms even without the photothermal feedback used in one of our earlier works \cite{merkel_coherent_2020}.

\subsection{Optical setup}
\begin{figure}[][h!]
\includegraphics[width=\columnwidth]{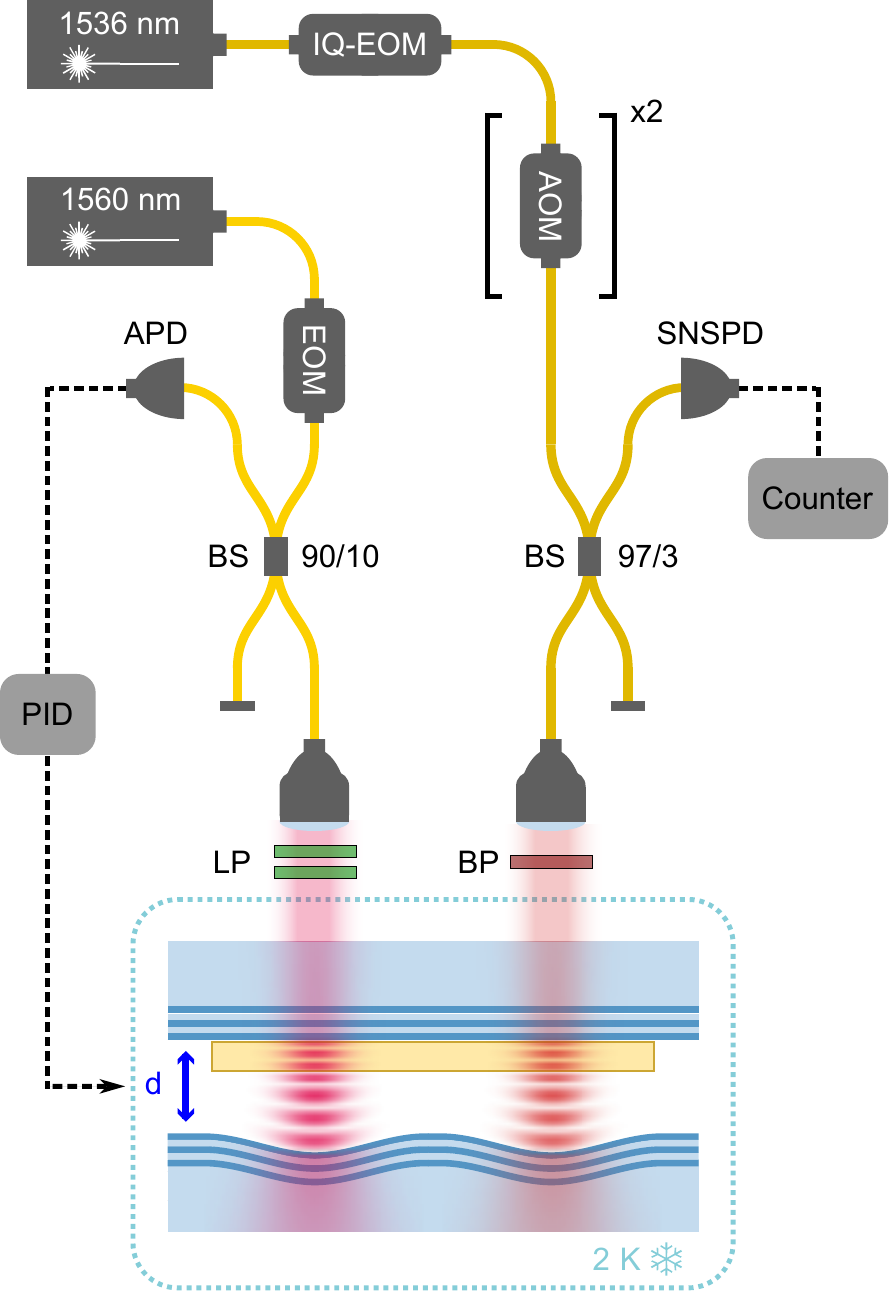}
\caption{\label{fig:SetupDetail} Experimental setup. The system uses two cavities fabricated on the same pair of mirrors. The first resonator (left half of the image) is used to stabilize the mirror separation using the Pound–Drever–Hall technique. The necessary sidebands at $f=f_0\pm\SI{280}{\mega\hertz}$ are generated using an electro-optic modulator (EOM). Using a beamsplitter (BS) with $90/10$ split ratio, the backreflection of the cavity is directed to an avalanche photodiode (APD). The mode is passing through two longpass filters (LP) with a cut-off at \SI{1550}{\nano\meter} to suppress amplified spontaneous emission noise.\\ In the second cavity (right half), individual erbium emitters are investigated and controlled. To this end, frequency modulation of the light is performed using an IQ-EOM and two consecutive acousto-optical modulators (AOM). A beamsplitter with a $97/3$ splitting ratio directs the majority of the light emitted from the cavity to a superconducting nanowire single-photon detector (SNSPD). The mode is passing through a bandpass filter (BP) with a center wavelength of \SI{1536}{\nano\meter} and a pass-band of \SI{3}{\nano\meter} to avoid detecting scattered light from the lock laser and other light sources.}
\end{figure}

The optical setup is shown in Figure~\ref{fig:SetupDetail}. It has two separate branches, one for stabilizing the cavity mirror separation, and the other for coherent manipulation of the erbium emitters and the detection of their emission. For the first task, a tuneable diode laser (Toptica CTL) is used, which is locked to a frequency comb with a linewidth of $\sim \SI{0.2}{\mega\hertz}$ (Menlo Systems FC1500). For coherent excitation, a fiber laser is used (Koheras Basik X15), which exhibits a better short-term-stability ($\lesssim \SI{1}{\kilo\hertz}$). To ensure long-term stability, the laser is again stabilized to the frequency comb and referenced to a wavemeter (HighFinesse WS8).

The light that excites the dopants is modulated by an I/Q electro-optical modulator (IxBlue MXIQ-LN-30) that allows for frequency shifts of several \si{\giga\hertz}. In addition, two acousto-optical modulators (Gooch and Housego FiberQ) are used for narrow-bandwidth frequency modulation, phase modulation, and the temporal shaping of the laser pulses. In addition, they allow for switching with an on-off contrast of 120~dB to avoid unwanted leakage of the pump laser light. The electronic signals to control the experiment are generated with arbitrary waveform generators (Zurich Instruments HDAWG and SHFSG).

The erbium emission is detected with superconducting nanowire single photon detectors (ID Quantique) with a detection efficiency of up to $80\,\%$. We choose a lower bias current to reduce the dark count rate to $\SI{10.2}{\hertz}$, which, however, also reduces the efficiency to  $40\,\%$. To avoid latching the detectors by the strong excitation pulses, an additional switch (PhotonWares NanoSpeed) is used in front of the detectors. The signal is then recorded using a time-tagger (Swabian Instruments Time Tagger 20). Spectral filters (Semrock) are used to eliminate the detrimental effect of stray light.

\medskip
\textbf{Funding} \par
This project received funding from the European Research Council (ERC) under the European Union's Horizon 2020 research and innovation program (grant agreement No 757772), from the Deutsche Forschungsgemeinschaft (DFG, German Research Foundation) under the German Universities Excellence Initiative - EXC-2111 - 390814868, from the German Federal Ministry of Education and Research (BMBF) via the grant agreement No 16KISQ046, and from the Munich Quantum Valley, which is supported by the Bavarian state government with funds from the Hightech Agenda Bayern Plus.

\medskip
\textbf{Conflicts of interest} \par 
The authors declare no conflicts of interest.

\medskip
\textbf{Data availability} \par
The datasets generated in this study are available via the following DOI: 10.14459/2024mp1733455.

\medskip

\bibliography{bib2}

\end{document}